\documentclass{amsart}
\begin{document}

\newtheorem{theorem}{Theorem}[section]
\newtheorem{defi}{Definition}[section]
\newtheorem{lemma}{Lemma}[section]
\newtheorem{prop}{Proposition}[section]
\newtheorem{coro}{Corollary}[section]

\newtheorem{exampleHLP}{Example}[section]
\newenvironment{example}[0]{\begin{exampleHLP}\rm}{\end{exampleHLP}}

\newtheorem{remarkHLP}{Remark}[section]
\newenvironment{remark}[0]{\begin{remarkHLP}\rm}{\end{remarkHLP}}
\def\Tr {{\rm Tr}\,}
\def\Md {\mathcal M}
\def\Pd {\mathcal P}
\def\Xd {\mathcal X}
\def\Yd {\mathcal Y}
\def\Zd {\mathcal Z}
\def\Wd {\mathcal W}
\title{Flat connections and Wigner-Yanase-Dyson metrics}
\author{Anna Jen\v cov\'a}
\address{Mathematical Institute, Slovak Academy of Sciences,
      \v Stef\'anikova 49 SK-814 73 Bratislava, Slovakia}
\email{jenca@mat.savba.sk}
\maketitle

{\bf Abstract.}
{\small On the manifold of positive definite matrices, we investigate the
existence of pairs of flat affine connections, dual with respect to a given
monotone metric. The connections are defined either using the
$\alpha$-embeddings and finding the duals with respect to the metric,
or by means of contrast functionals. We show that in both cases,
 the existence of such a
pair of connections is possible if and only  if the metric is given by the
Wigner-Yanase-Dyson skew information.}


\section{Introduction.}
An important feature of the classical information geometry is the
uniqueness of its structures,  the Fisher metric and the family of
affine $\alpha$-connections on a manifold ${\Pd}$ of probability distributions,
\cite{chentsov, amari}. In case of
finite quantum systems, this uniqueness does not take place: it
was shown by Chentsov and Morozova \cite{chemo} and later by Petz
\cite{petz96} that there are infinitely many Riemannian metrics,
that are monotone with respect to stochastic maps. As for the
affine connections, there were several definitions of the
$\alpha$-connections, \cite{ja,naga,hase97,hase02}.

In commutative case, two equivalent definitions of
the connections were used by Amari \cite{amari}. First, the connections can be
defined using $\alpha$-embeddings ($\alpha$-representations) given
by  the family of functions
\begin{equation}\label{eq:falfa}
f_{\alpha}(x)=\left\{\begin{array}{lc}
\frac2{1-\alpha}x^{\frac{1-\alpha}2},
                                             & \alpha\neq 1\\
                                        \log(x), & \alpha=1
                      \end{array}\right.
\end{equation}
On the other hand, the connections can be defined as  mixtures
of the exponential and the mixture connections,
\begin{equation}\label{eq:mixtures}
\nabla^{(\alpha)}=\frac{1+\alpha}2\nabla^{(e)}+\frac{1-\alpha}2\nabla^{(m)}
\end{equation}
Such connections are torsion-free and the $\alpha$ and $-\alpha$ connections are dual with
respect to the Fisher metric. Moreover,
in case of a finite system, that is on the manifold of all (non-normalized)
multinomial distributions, the $\alpha$-connections are flat for
all $\alpha$.

The definition involving  $\alpha$-representations can be easily generalized to
non-commutative case to obtain a family of flat connections $\nabla^{(\alpha)}$
on the manifold of
positive definite matrices. This definition was treated also
by the present author in \cite{ja} and \cite{ja2}.
The dual of such $\alpha$-connection with respect
to a given monotone metric
is in general different from the $-\alpha$-connection.
The duals  have vanishing Riemannian curvature,
but are not always torsion-free and hence not flat.
 The condition that the dual of the
$\alpha$-connection with respect to a monotone  metric is torsion-free
restricts $\alpha$ to the interval $[-3,3]$ and, for such $\alpha$,
 singles out a monotone metric
$\lambda_{\alpha}$, which belongs to the family of Wigner-Yanase-Dyson (WYD)
metrics.
This is also equivalent to the condition that the dual of $\nabla^{(\alpha)}$
is $\nabla^{(-\alpha)}$, see also \cite{grasselli}.  A brief description of these results is
given in sections \ref{section:metric} and
\ref{section:affine}.

For $\alpha=\pm1$, we get the BKM metric, with
respect to which the mixture $\nabla^{(m)}$ and exponential $\nabla^{(e)}$
connections are dual. As in the classical case,
we may use mixtures   of $\nabla^{(e)}$ and
$\nabla^{(m)}$ to
define a family of torsion-free connections, having the required duality
properties with respect to the BKM metric. In our approach, however,
the value of $\alpha$ in (\ref{eq:mixtures}) will be restricted to
the interval $[-1,1]$, but the proofs in Section
\ref{section:pconnections} suggest that our results hold more
generally. Convex mixtures were
considered also by Grasselli and Streater, see the Discussion in  \cite{grastre}.
We will answer the questions discussed there in proving that, for
$\alpha \in (-1,1)$, affine connections defined by (\ref{eq:mixtures})
are different from the $\alpha$-connections and are not flat.
A simple direct
proof of this fact can be found at the end of Section
\ref{section:pconnections}.

Another way to define an affine connection was proposed by Eguchi in
\cite{eguc83}, by means of
a  contrast functional on $\Pd$. A functional $\phi:\Pd\times \Pd\to \mathbb
R$ is a contrast functional if it
 satisfies $\phi(p,q)\geq0$ for all $p,q$ and $\phi(p,q)=0$ if and only if
$p=q$.  Using
such a functional, a metric tensor and affine connection can be defined: Let
$\theta_1,\dots,\theta_p$ be a smooth parametrization of $\Pd$ and let
$\partial_i$, $i=1,\dots,p$ be the corresponding vector fields, then the
metric tensor is given by
$$
g^{\phi}_{ij}=-\partial_i\partial'_j\phi(p(\theta),p(\theta'))|_{\theta=\theta'}
$$
An affine connection $\nabla^{\phi}$ is defined by
$$
\Gamma^{\phi}_{ijk}=g^{\phi}(\nabla^{\phi}_{\partial_i}\partial_j,\partial_k)=
-\partial_i\partial_j\partial_k'\phi(p(\theta),p(\theta'))|_{\theta=\theta'}
$$

Consider a special class of contrast functionals $\phi_g$, related to  convex
functions $g$ satisfying $g(1)=0$ by
$$
\phi_g(p,q)=\int g(\frac qp)dp
$$
In this case, it was shown \cite{amari} that the corresponding metric is
the Fisher metric (multiplied by $g''(1)$)
and the affine connection is the $\alpha$-connection,
$\alpha=2g'''(1)+3$. As the quantum counterpart of such functionals, we will
use the relative $g$-entropies $H_g$ defined by Petz \cite{petz86}
$$
H_g(\rho,\sigma)=\Tr \rho^{1/2}g(L_{\sigma}/R_{\rho})(\rho^{1/2})
$$
where $g$ is an operator convex function and $g(1)=0$.
It was shown that
\begin{enumerate}
\item[(a)] In the normalized case (or if $\Tr \rho=\Tr \sigma$),
$H_g(\rho,\sigma)\geq 0$ and $H_g(\rho,\sigma)=0$ if and only if
$\rho=\sigma$
\item[(b)] $H_g(\lambda\rho,\lambda\sigma)=\lambda H_g(\rho,\sigma)$ for each
$\lambda>0$,
\item[(c)] $H_g$ is jointly convex in $\rho$ and $\sigma$.
\item[(d)]
$H_g$ is monotone, that is, it decreases under stochastic maps,
\item[(e)] $H_g$ is differentiable.
\end{enumerate}

We see that $H_g$ is a contrast functional on the manifold of
quantum states, and we will show that we can use it to define the
geometrical structures as above even in the non-normalized case.
The relative $g$-entropies were used by Lesniewski and Ruskai
\cite{leru99}, who proved that the Riemannian structure, given by
$H_g$ is monotone for each $g$  and, conversely, each monotone
metric is obtained in this way. A short account on some of their
results is in section \ref{section:entropy}.

In section \ref{section:pconnections}, we will use $H_g$ to define an
affine connection and show that this definition contains both the
$\alpha$-connections, defined from $\alpha$-embeddings, and the convex
mixtures of $\nabla^{(m)}$ and $\nabla^{(e)}$. We will show that for each
monotone metric, there is a family of such connections
(the $p$-connections), parametrized by
$p\in[0,1]$, such that these are torsion-free and the $p$- and
$(1-p)$- connections are dual. We will then use the theory of statistical
manifolds by Lauritzen \cite{abklr} to investigate the Riemannian curvature
of the connections.

Finally, in the last section we will show that  a pair of dual
flat connections exists if and only if the metric is one of the
WYD metrics $\lambda_{\alpha}$. The flat connections are then the
$\pm \alpha$-connections. This result holds  for the
connections given by the relative $g$-entropies. It is known from
\cite{amari} that dual
 flat connections give rise
to divergence functionals on the manifold, it is therefore reasonable to
consider connections defined from functionals having the properties (a)-(e).
The class of $g$-entropies seems to be large enough, although it
does not contain all such functionals (see \cite{leru99}).
The main results of the present paper can be summarized as
follows: If a pair of dual flat connections is required, the
structures of information geometry are unique even in the quantum case, at
least if we consider only connections defined by the relative $g$-entropies.
 These structures are provided by the family of
Wigner-Yanase-Dyson metrics and the $\alpha$-connections.

\section{The manifold and monotone metrics.}\label{section:metric}

Let $\Md_n(\mathbb C)$ be the space of $n\times n$ complex
matrices, $\Md_h$ be the real linear subspace of hermitian matrices
and let $\Md\subset \Md_h$ denote the set of positive
definite matrices. As an open subset in a finite dimensional real vector space,
$\Md$ inherits the structure of a differentiable manifold. The
tangent space $T_{\rho}$ of $\Md$ at $\rho$ is the linear space of
directional (Fr\`echet) derivatives in the direction of smooth
curves in $\Md$ and it can be identified with $\Md_h$ in an obvious
way. In the present paper, the elements of the tangent space, seen as
 directional derivative operators, will be denoted by $\Xd,\Yd$,
 etc., while the corresponding capital letters will mean their representations
 $X=\Xd(\rho)$ etc.  in  $\Md_h$. The map $\Xd\mapsto X$
is the same as Amari's -1-representation of the tangent space in
the classical case  \cite{amari}, see also the next Section. The
vector fields on $\Md$ are represented by
$\Md_h$-valued functions on $\Md$. If ${\Xd},{\Yd}$ are
vector fields, then the bracket $[{\Xd},{\Yd}]$ is
unrelated to the usual commutator of the representing matrices and
these two should not be confused. In the present paper, we will
use $[\cdot,\cdot]$ only in the first (vector fields) meaning.

A Riemannian structure is introduced in $\Md$ by
$$
\lambda_{\rho}(X,Y)=\Tr XJ_{\rho}(Y),\quad  X,Y\in T_{\rho}
$$
where $J_{\rho}$ is a suitable operator on matrices. We say that the metric
$\lambda$ is monotone if it is monotone with respect to stochastic maps,
that is, we have
$$
\lambda_{T(\rho)}(T(X),T(X))\leq \lambda_{\rho}(X,X),\ \rho\in \Md,\ X\in
T_{\rho}
$$
for a stochastic map $T$. It is an important result of Petz \cite{petz96},
that this is equivalent to
$$
J_{\rho}=R_{\rho}^{-1/2}F(L_{\rho}/R_{\rho})^{-1}R_{\rho}^{-1/2}
$$
where $F:\mathbb R^+\to \mathbb R$ is an operator monotone function,
which is symmetric, $F(x)=xF(x^{-1})$, and normalized, $F(1)=1$.
The operators $L_{\rho}$ and $R_{\rho}$ are the left and right
multiplication operators. Clearly, $J_{\rho}(X)=\rho^{-1}X$ if $X$ and $\rho$
commute, so that the restriction of $\lambda$ to commutative submanifolds is
the Fisher metric.

\begin{example} Let $J_{\rho}$ be the symmetric logarithmic
derivative, given by $J_{\rho}(X)=Y$, $Y \rho+\rho Y=2X$, then the metric
$\lambda$ is monotone, with $F(x)=\frac{1+x}2$. This metric is sometimes
called the Bures metric and it is the smallest monotone Riemannian metric.
\end{example}

\begin{example} The largest monotone metric is given by the
operator monotone function $F(x)=\frac{2x}{1+x}$. In this case,
$J_{\rho}(X)=
\frac12(\rho^{-1}X+X\rho^{-1})$ is the right logarithmic derivative (RLD).
\end{example}

\begin{example} An important example of a monotone metric is the
Kubo-Mori-Bogoljubov (BKM) metric, obtained from
$$
\frac{\partial^2}{\partial s\partial t}\Tr (\rho+sX)\log(\rho+tY)|_{s,t=0}
=\lambda_{\rho}(X,Y)
$$
In this case, $F(x)=\frac{x-1}{\log(x)}$.
\end{example}

\section{The $\alpha$-representation and
$\alpha$-connections.}\label{section:affine}

Let $f:\mathbb R\to \mathbb R$ be a monotone function and let
$\rho\in \Md$. Let us define the operator $L_f[\rho]:\ \Md_h\to \Md_h$ by
$$
L_{f}[\rho](X)=\frac d{ds}f(\rho+sX)|_{s=0}
$$
This operator has the following properties \cite{ja}:
\begin{enumerate}
\item[(i)] The chain rule: $L_{f\circ g}[\rho]=L_f[g(\rho)]L_g[\rho]$. In particular,
if $f$ is invertible then $L_f[\rho]$ is invertible and
$L_f[\rho]^{-1}=L_{f^{-1}}[f(\rho)]$.
\item[(ii)] $L_f[\rho]$ is a self adjoint operator in $\Md_h$, with respect
to the
Hilbert-Schmidt inner product $\langle X,Y\rangle=\Tr X^*Y$.
\item[(iii)] If $X\rho=\rho X$, then $L_f[\rho](X)= f'(\rho)X$, $f'(x)=
\frac d{dx}f(x)$.
\end{enumerate}

Let now $f_{\alpha}$ be given by
(\ref{eq:falfa}). The map
$$
\ell_{\alpha}: \Md\ni \rho \mapsto f_{\alpha}(\rho)\in \Md_h
$$
will be called the $\alpha$-embedding of $\Md$.  The $\alpha$-embedding
induces the map
$$
T_{\rho}\ni X\mapsto \Xd(f_{\alpha}(\rho))=L_{\alpha}[\rho](X)\in \Md_h
$$
where $L_{\alpha}[\rho]:=L_{f_{\alpha}}[\rho]$, it will be called the
$\alpha$-representation of the tangent vector $X$.
We will often omit the indication of the point in the square brackets, if no
confusion is possible.

Let $\lambda$ be a monotone metric
and let $Y_1=L_{\alpha}(X_1)$ and $Y_2=L_{\alpha}(X_2)$ be the
$\alpha$-representations of the tangent vectors $X_1$ and $X_2$, then
\begin{equation}\label{eq:alphrep}
\lambda_{\rho}(X_1,X_2)=\Tr Y_1 K_{\alpha}(Y_2)
\end{equation}
where $K_{\alpha}=L_{\alpha}^{-1}J_{\rho}L_{\alpha}^{-1}$.

\begin{example} The family of Wigner-Yanase-Dyson (WYD) metrics
$\lambda_{\alpha}$ is defined by $J_{\rho}=L_{-\alpha}L_{\alpha}$.
In \cite{hape}, it was shown  that such metrics are monotone for
$\alpha\in [-3,3]$ and that there are no other monotone metrics,
satisfying
$$
\lambda_{\rho}(X,Y)=\frac{\partial^2}{\partial s\partial t}\Tr
f(\rho+sX)f^*(\rho+tY)|_{s,t=0}
$$
for some functions $f$ and $f^*$. The corresponding operator
monotone function is
$$
F_{\alpha}(x)=\frac{1-\alpha^2}4\frac{(x-1)^2}{(x^{\frac{1+\alpha}2}-1)(
x^{\frac{1-\alpha}2}-1)}
$$
As special cases, we obtain the BKM metric for $\alpha=\pm 1$ and
RLD metric for $\alpha=\pm 3$. The smallest metric in this class is
the Wigner-Yanase (WY) metric, corresponding to $\alpha=0$, here
$F_0(x)=\frac14(1+\sqrt(x))^2$, the Bures metric is not included.
For the metric $\lambda_{\alpha}$,
$\alpha\in [-3,3]$, we have $K_{\alpha}=L_{-\alpha}L_{\alpha}^{-1}$.
It can be shown that  $K_{\alpha}^{-1}=K_{-\alpha}$  if and only if
$\lambda=\lambda_{\alpha}$.
\end{example}

The connection $\nabla^{(\alpha)}$ is defined by
$$
L_{\alpha}((\nabla^{(\alpha)}_{\Xd}{\Yd})(\rho))
={\Xd}{\Yd} f_{\alpha}(\rho)
$$
for smooth vector fields $\Xd,\Yd$.
Clearly,  a vector field  is parallel with respect to this
connection if and only if its $\alpha$-representation is a constant
hermitian matrix valued function on $\Md$.
For $\alpha=-1$ and $\alpha=1$, we get the mixture and
exponential connections, sometimes denoted by $\nabla^{(m)}$ and
$\nabla^{(e)}$. The mixture connection coincides with the natural
flat affine structure inherited from $\Md_h$.

For each $\alpha$, there is a coordinate system
$\xi_1,\dots,\xi_N$, such that
$f_{\alpha}(\rho(\xi))=\sum_i\xi_i Z_i$, where $Z_i\in \Md_h$,
$i=1,\dots,N$ form a basis of $\Md_h$. Clearly, such coordinate system is
$\nabla^{(\alpha)}$-affine. The existence of an affine coordinate
system is equivalent to flatness of the connection
$\nabla^{(\alpha)}$, that is, the connections are torsion-free and
the Riemannian curvature tensor vanishes. Thus we have a
one-parameter family of flat $\alpha$-connections, just as in the
classical case. But, contrary to the classical case, the
$\nabla^{(\alpha)}$ and $\nabla^{(-\alpha)}$ are not dual for a
general monotone metric.

Let us define the connection $\nabla^{(\alpha)*}$ by
$$
L_{\alpha}^{-1}J_{\rho}((\nabla^{(\alpha)*}_{\Xd}{\Yd})(\rho))
={\Xd}L_{\alpha}^{-1}J_{\rho}(Y)=
{\Xd}K_{\alpha}L_{\alpha}(Y)
$$

It can be easily seen from (\ref{eq:alphrep}) that the connections
$\nabla^{(\alpha)}$ and $\nabla^{(\alpha)*}$ are dual with respect
to $\lambda$. It follows that $\nabla^{(\alpha)*}$ is also
curvature free  and it is torsion-free if and only if \cite{ja}
\begin{equation}\label{eq:torsion}
{\Xd} L_{\alpha}^{-1}J_{\rho}(Y)={\Yd}
L_{\alpha}^{-1}J_{\rho}(X)
\end{equation}
for all vector fields satisfying $[{\Xd},{\Yd}]=0$.

\begin{theorem}\label{theorem:torsion}\cite{ja2}
Let $\alpha\in [-3,3]$. The following are equivalent.
\begin{enumerate}
\item[(i)] $(\nabla^{(\alpha)})^*$  is torsion-free
\item[(ii)] $J_{\rho}=L_{\alpha}L_{-\alpha}$
\item[(iii)] $(\nabla^{(\alpha)})^*=\nabla^{(-\alpha)}$
\end{enumerate}
 \end{theorem}

\begin{proof} (i)$\implies$(ii):\  Let $\theta\mapsto \rho(\theta)$ be a
smooth parametrization of $\Md$ and let
$\partial_i=\frac{\partial}{\partial \theta_i}$, $i=1,\dots, N$.
Let us denote
$X_i(\theta)=\partial_i(\rho(\theta))$.
Let $\nabla^{(\alpha)*}$ be torsion-free
and let $F_i(\theta)=L_{\alpha}^{-1}J_{\rho(\theta)}(X_i(\theta))$,
$i=1,\dots, N$. Then  we get from (\ref{eq:torsion}) that
$\partial_jF_i=\partial_iF_j$ for all $i,j$.

Let $A_1,\dots,A_N$ be a basis
of $\Md_h$ and let $F_i(\theta)=\sum_kf_{ik}(\theta)A_k$, then $\partial_i
f_{jk}(\theta)=\partial_j f_{ik}(\theta)$ for all $k$, $i$ and $j$.
This implies the
existence of functions $\phi_1,\dots,\phi_N$, such that
$f_{ik}(\theta)=\partial_i
\phi_k(\theta)$. Let $\phi(\theta)=\sum_k\phi_k(\theta)A_k$, then
$F_i=\partial_i\phi$. Moreover, if $\rho_t=\rho(\theta(t))$ is a curve in
$\Md$, then
$$
\frac d{dt}\phi(\theta(t))=\sum_i\frac d{dt}\theta_i(t)F_i(\theta(t))
=L_{\alpha}^{-1}[\rho_t]
J_{\rho_t}(\frac d{dt}\rho_t)
$$
Let now $\rho\in \Md$ and let us consider the curve
$\rho_t=\rho(\theta(t))=t\rho+(1-t)$. Using the fact that
 $\frac d{dt}\rho_t=\rho-1$
and $\rho_t$ commute for all $t$,  we have
\begin{eqnarray*}
\phi(\theta(1))-\phi(\theta(0))&=&\int_0^1\frac d{dt}\phi(\theta(t))dt=
\int_0^1 L_{\alpha}^{-1}[\rho_t]J_{\rho_t}(\rho-1)dt=\\
&=&\int_0^1(1+t(\rho-1))^{\frac{\alpha-1}2}(\rho-1)dt=f_{-\alpha}(\rho)-f_{-\alpha}(I)
\end{eqnarray*}
Therefore, $\phi(\theta)=f_{-\alpha}(\rho(\theta))+c$. It follows that
$$
L_{\alpha}^{-1}J_{\rho(\theta)}(X_i(\theta))=F_i(\theta)=\partial_i
f_{-\alpha}(\rho(\theta))=L_{-\alpha}(X_i(\theta))
$$
and $J_{\rho}=L_{\alpha}L_{-\alpha}$.

(ii)$\implies$(iii) and (iii)$\implies$(i) are quite clear.
\end{proof}

The statement for $\alpha=\pm 1$ was already proved in
\cite{amana}. The equivalence (ii)$\iff$(iii) was proved (by a
different method) in \cite{grastre} for $\alpha=\pm 1$ and
\cite{grasselli} for $\alpha\in (-1,1)$.

\begin{remark} Let ${\mathcal D} =\{ \rho \in \Md\ :\ \Tr \rho =1\}$ be the
 submanifold of quantum states.
The connections induced on $\mathcal D$ are orthogonal projections of
the above connections. The Riemannian
curvature is given by \cite{ja}
$$
R^{\alpha}_{\rho}(X,Y,Z,W)=\frac{1-\alpha^2}4\{ \Tr YJ_{\alpha}(Z)
\Tr XJ_{\rho}(W)-\Tr XJ_{\alpha}(Z)\Tr YJ_{\rho}(W)\}
$$
where $\rho\in {\mathcal D}$, $X,Y,Z,W\in T_{\rho}({\mathcal D})$, and
thus $R^{\alpha}=0$ if and only if $\alpha=\pm 1$. Therefore, the
$\alpha$-connections are not flat on $\mathcal D$, unless $\alpha=\pm1$,
which corresponds to the classical results.
\end{remark}

\section{Relative $g$-entropies and monotone metrics.}\label{section:entropy}

Let $G$ be the set of all operator convex functions $(0,\infty)\to \mathbb R$,
satisfying $g(1)=0$ and $g''(1)=1$. For $g\in G$, we define the relative
$g$-entropy $H_g:\ {\mathcal M}\times {\mathcal M}\to \mathbb R$ by
\cite{petz86}
$$
H_g(\rho,\sigma)=\Tr \rho^{1/2}g(L_{\sigma}/R_{\rho})(\rho^{1/2})
$$
The set  $G$ is the set of functions of the form
\begin{equation}\label{eq:G}
g(u)=a(u-1)+\int_{[0,\infty]}(u-1)^2\frac{1+s}{u+s}d\mu(s)
\end{equation}
where $\mu$ is a positive finite measure on $[0,\infty]$ satisfying
$\int_{[0,\infty]}d\mu(s)=1/2$ and $a=g'(1)$ is a real number. We will denote
$b=\mu(\{\infty\})$ and $c=\mu(\{0\})$
the possible atoms in $0$ and $\infty$, then
$$
g(u)=a(u-1)+b(u-1)^2+c\frac{(u-1)^2}u+\int_0^{\infty}(u-1)^2\frac{1+s}{u+s}d\mu(s)
$$

For an operator convex function $g$ we define its transpose $\hat
g(u)=ug(u^{-1})$. Clearly,  $g\in G$ implies $\hat g\in G$, with the
positive
measure $\hat \mu$ satisfying $d\hat \mu(s)=d\mu(s^{-1})$ and $\hat a=-a$.
We
say that $g$ is symmetric if $g=\hat g$. For each symmetric function
$h\in G$,
we denote by $G_h\subset G$ the convex subset of functions such that $g+\hat
g=2h$. If $g\in G_h$, then clearly $\hat g \in G_h$ and $H_{\hat g}(\rho,
\sigma)=H_g(\sigma,\rho)$.

\begin{theorem} \cite{leru99} \label{theorem:Hg} For each $\rho,\sigma\in \Md$,
\begin{eqnarray*}
H_g(\rho,\sigma)&=&a\Tr(\sigma-\rho)+\\
&+&\Tr(\sigma-\rho)\{
b\rho^{-1}+c\sigma^{-1}+\int_0^{\infty}\frac{1+s}{L_{\sigma}+sR_{\rho}}d\mu(s)\}(\sigma-\rho)\\
&=&
a\Tr(\sigma-\rho)+\Tr(\sigma-\rho)R_{\rho}^{-1}k(L_{\sigma}/R_{\rho})(\sigma-\rho)
\end{eqnarray*}
where
\begin{equation}\label{eq:k}
k(u)=\int_{[0,\infty]}\frac{1+s}{u+s}d\mu(s)=\frac{g(u)-a(u-1)}{(u-1)^2}
\end{equation}
\end{theorem}

The above Theorem implies that if $a=0$, $H_g$
is a contrast functional on $\Md$.
The value of $g'(1)=a$ does not influence the Riemannian structure and
connections defined by $H_g$, so that we may also use functions with
$g'(1)\neq 0$, as it is sometimes more convenient, for example,
$g(u)=-\log u$.

Let us consider the mixture connection $\nabla^{(m)}$ on $\Md$. A
vector field on $\Md$ is  parallel with respect to $\nabla^{(m)}$
if and only if its -1-representation is  a constant $\Md_h$-valued
function over $\Md$.
In the
rest of the paper, we will deal only with such vector fields. The symbol $\Xd$
will denote the vector field such that the constant value of the -1-representation
is $X$, similarly $\Yd$, etc.
Note that for such vector fields, we
have $[{\Xd},{\Yd}]=0$.

Let us define the Riemannian metric $\lambda^g$ on $\Md$ by
\begin{equation}\label{metric}
\lambda^g_{\rho}(X,Y)=
-\frac{\partial^2}{\partial s\partial t}H_g(\rho+sX,\rho+tY)|_{s,t=0},
\quad \forall X,Y \in T_{\rho}
\end{equation}
Then \cite{leru99}
$$
\lambda^g_{\rho}( X,Y)=
\Tr X R_{\rho}^{-1}k_{sym}(L_{\rho}/R_{\rho})(Y)
$$
with
$$k_{sym}(u)=k(u)+u^{-1}k(u^{-1})=\frac{g(u)+\hat g(u)}{(u-1)^2}$$
Moreover, the function $k_{sym}$ is operator monotone decreasing, hence
$\lambda^g$ is a monotone metric, with $F=1/{k_{sym}}$ the corresponding
operator monotone function. Note also that if $h$ is a fixed symmetric
function in $G$, then $\lambda^g$ defines the same monotone metric for
each $g\in G_h$.

Conversely, if $\lambda$ is a monotone metric with
the operator monotone function $F$, then
\begin{equation}\label{eq:h}
h(u)=\frac12\frac{(u-1)^2}{F(u)}
\end{equation}
is a symmetric operator convex function with $h(1)=0$, so that
$\lambda=\lambda^h$. The condition $h''(1)=1$ is
equivalent to the normalization condition $F(1)=1$.
This gives a one-to-one correspondence
between the monotone metrics and the convex sets $G_h$, with symmetric
$h\in G$.

\section{The $p$-connections.}\label{section:pconnections}

Let us fix a monotone metric $\lambda$ and let
$h$ be given by (\ref{eq:h}).
Let us choose some $g\in G_h$, then $\lambda=\lambda^g$. We define the
affine connection $\nabla^{(g)}$ by
$$
\lambda_{\rho}(\nabla^{(g)}_{\Xd}\Yd,\Zd)=
-\frac{\partial^3}{\partial s\partial t\partial
u}H_g(\rho+sX+tY,\rho+uZ)|_{s,t,u=0}
$$
just as in the classical case. It is clear that the restriction of
$\nabla^{(g)}$ to  submanifolds of mutually commuting elements coincides
with the classical $\alpha$-connection, with $\alpha=2g'''(1)+3$.
In contrast with the classical case, the condition $g\in G$
leads to a restriction on $\alpha$. Indeed, we have
$$
g'''(1)= -6 \int_{[0,\infty]}\frac1{1+s}d\mu(s)
$$
From this, $0\geq g'''(1)\geq -3$ and therefore
$\alpha\in[-3,3]$ for each $g\in G$.

\begin{prop} The connections $\nabla^{(g)}$ and $\nabla^{{(\hat g)}}$ are dual with
respect to $\lambda$. Moreover, the connections are torsion-free.
\end{prop}

\begin{proof} We have
\begin{eqnarray*}
{\Xd}\lambda_{\rho}(\Yd,\Zd)&=&
-\frac d{du}\frac{\partial^2}{\partial t\partial s}
H_g(\rho+uX+sY,\rho+uX+tZ)|_{s,t,u=0}\\
&=&-\frac {\partial^3}{\partial s\partial
t\partial u}H_g(\rho+uX+sY,\rho+tZ)|_{s,t,u=0}-\\
&-&\frac{\partial^3}{\partial
s\partial t\partial u}H_{\hat g}(\rho+uX+tZ,\rho+sY)|_{s,t,u=0}=\\
&=&\lambda_{\rho}(\nabla^{(g)}_{\Xd}{\Yd},{\Zd})+
\lambda_{\rho}(\Yd,\nabla^{(\hat g)}_{\Xd}\Zd)
\end{eqnarray*}
so that duality is proved. Moreover, as $[\Xd,\Yd]=0$,
the connection is torsion-free if $\nabla^{(g)}_{\Xd}\Yd-
\nabla^{(g)}_{\Yd}\Xd=0$, which is obvious.
\end{proof}

If the function $g$ is symmetric, then from the previous Proposition,
$\nabla^{(g)}$ is
self-dual and torsion-free, hence it is the metric connection $\bar{\nabla}$.
For $g\neq\hat g$, let us  define
$g_p=pg+(1-p)\hat g$, then $g_p\in G_h$ for  $p\in[0,1]$
and $\hat g_p=g_{1-p}$. For $\lambda$ and $g$ fixed,  the
connection given by $g_p$ will be called the $p$-connection and denoted by
$\nabla^{(p)}$. Clearly, $\nabla^{(p)}$ is a convex mixture of
$\nabla^{(g)}$ and $\nabla^{(\hat g)}$,
$$
\nabla^{(p)}=p\nabla^{(g)} + (1-p)\nabla^{(\hat g)}
$$
Thus we have a one-parameter family of torsion-free $p$-connections, satisfying
$(\nabla^{(p)})^*=\nabla^{(1-p)}$. We have $\nabla^{(1/2)}=\bar {\nabla}$ for all
$g\in G_h$. In the rest of this Secion, we will investigate the Riemannian
curvature of the $p$-connections.

\begin{example} We see from (\ref{eq:G}) that the extreme boundary of $G$
consists of functions
\begin{eqnarray*}g_s(u)&=&\frac{1+s}2\frac{(u-1)^2}{u+s}\ \mbox{for s}
\geq 0\\
g_{\infty}(u)&=&\frac12 (u-1)^2
\end{eqnarray*}
We have $\hat g_s=g_{s^{-1}}$ for $s>0$ and $\hat g_0=g_{\infty}$.
In this case
$$
G_{h_s}=\{g_p=pg_s+(1-p)\hat g_s,\ p\in [0,1]\}
$$
where $h_s=\frac12(g_s+\hat g_s)$. For the corresponding metric  we obtain a
unique family of $p$-connections. In particular, if $s=1$,
$g_1=h_1$ is symmetric and $G_{h_1}=\{h_1\}$. The corresponding metric is the
Bures metric. Hence we see that for the Bures metric,
we obtain only the metric connection,
which is known to be not flat, see for example \cite{ditt99}.
\end{example}

\begin{example}Let
$$
g_{\alpha}(u)=\left\{ \begin{array}{lc}
\frac4{1-\alpha^2}(\frac{1+u}2-u^{\frac{1+\alpha}2})& \alpha\neq \pm1\\
-\log u& \alpha=-1\\
u\log u& \alpha=+1
\end{array}\right.
$$
Then $g_{\alpha}\in G$ for $\alpha\in[-3,3]$ and
$\hat g_{\alpha}=g_{-\alpha}$.
The relative entropies $H_{g_{\alpha}}$ are (up to a linear term) the
$\alpha$-divergences defined by Hasegawa in \cite{hase93}. It was also proved
that $\lambda^{g_{\alpha}}=\lambda_{\alpha}$, the WYD metric, and
$\nabla^{g_{\alpha}}=\nabla^{(\alpha)}$, the $\alpha$-connection
from Section \ref{section:affine}, see
also \cite{hase02}.  Hence, $\nabla^{(g_{\alpha})}$ is flat. In
particular, for $\alpha=\pm1$ we get the BKM metric and the mixture and
exponential connection. The family of $p$-connection for $g(u)=-\log(u)$ is
$$
\nabla^{(p)}=p\nabla^{(m)}+(1-p)\nabla^{(e)}
$$
In the classical case, this is an equivalent definition of the
$\alpha$-connection, $p=(1-\alpha)/2$. In our case however, these connections
are different from the $\alpha$-connections, which, by Theorem
\ref{theorem:torsion}, have torsion-free duals with respect to the BKM metric
if and only if $\alpha=\pm1$.
\end{example}

To compute the
Riemannian curvature tensor of $\nabla^{(p)}$, we use the theory of statistical
manifolds due to Lauritzen, \cite{abklr}.
A statistical manifold is a triple $(M,\lambda,\tilde D)$, where $M$ is a
differentiable manifold, $\lambda$ is the metric tensor and $\tilde D$ is a
symmetric covariant 3-tensor called the skewness.

On $M$, a class of connections is introduced by
\begin{equation}\label{eq:pconnection}
\nabla^{(p)}_XY=\bar{\nabla}_XY-\frac{1-2p}2D(X,Y),
\end{equation}
where $X,Y$ are smooth vector fields,
$\bar{\nabla}$ is the metric connection and the tensor $D$ is given by
$\tilde D(X,Y,Z)=\lambda(D(X,Y),Z)$. Such connections are torsion-free, this is
equivalent to symmetry of $\tilde D$ resp. $D$.
Moreover, $(\nabla^{(p)})^*=\nabla^{(1-p)}$. Let $R^{p}$ be the
corresponding Riemannian curvature. The manifolds satisfying
$R^p=R^{1-p}$ for all $p$ are called conjugate symmetric. It was proved in
\cite{abklr} that the manifold is conjugate symmetric if and only if the tensor
$F= \bar {\nabla}\tilde D$ is symmetric. From symmetry of $\tilde D$, it
follows that $F$ is symmetric if (and only if) it is symmetric in $X$ and $Y$.
We also have  that if there is some $p\neq 1/2$, such that
$R^{p}=R^{1-p}$, then the manifold is conjugate symmetric.

Let  $g\in G$, then  $(\Md,\lambda^g,\tilde D)$,
where $D(\Xd,\Yd)=\nabla^{(g)}_{\Xd}\Yd-\nabla^{(\hat g)}_{\Xd}\Yd$, is a
statistical manifold. The  connections defined by
(\ref{eq:pconnection}) coincide with the $p$-connections if $p\in [0,1]$.
For simplicity, we denote this manifold by
$(\Md, g)$. If $g$ is symmetric, then $\tilde D\equiv 0$ and
$\nabla^{(p)}=\bar{\nabla}$ for all $p$,
in this case, the manifold is trivially conjugate symmetric.

\begin{prop}\label{prop:Rp} Let $\bar R=R^{1/2}$. Then
\begin{eqnarray*}
R^{p}(\Xd,\Yd,\Zd,\Wd)&=& \bar
R(\Xd,\Yd,\Zd,\Wd)+\frac{1-2p}2\{F(\Yd,\Xd,\Zd,\Wd)-F(\Xd,\Yd,\Zd,\Wd)\}\\
&+& \frac{(1-2p)^2}4
\{\lambda(D(\Xd,\Wd),D(\Yd,\Zd)) -\lambda(D(\Xd,\Zd),D(\Yd,\Wd))\}
\end{eqnarray*}

\end{prop}

\begin{proof} We have $[\Xd,\Yd]=0$ and therefore
$$
R^p(\Xd,\Yd,\Zd,\Wd)=\lambda(\nabla^{(p)}_{\Xd}\nabla^{(p)}_{\Yd}{\Zd}-
\nabla^{(p)}_{\Yd}\nabla^{(p)}_{\Xd}\Zd,\Wd).
$$
Let us now recall that
$$
F(\Xd,\Yd,\Zd,\Wd)=\Xd\tilde D(\Yd,\Zd,\Wd)-
\tilde D(\bar{\nabla}_{\Xd}\Yd,\Zd,\Wd)-\tilde
D(\Yd,\bar{\nabla}_{\Xd}\Zd,\Wd)-\tilde D(\Yd,\Zd,\bar{\nabla}_{\Xd}\Wd)
$$
From (\ref{eq:pconnection}) we get
\begin{eqnarray*}
\nabla^{(p)}_{\Xd}\nabla^{(p)}_{\Yd}\Zd&=&
\bar{\nabla}_{\Xd}\bar{\nabla}_{\Yd}\Zd
-\frac{1-2p}2\{\bar{\nabla}_{\Xd}D(\Yd,\Zd)+
D(\Xd,\bar{\nabla}_{\Yd}\Zd)\}+\\
&+&\frac{(1-2p)^2}4D(\Xd,D(\Yd,\Zd))
\end{eqnarray*}
Moreover, from self-duality of $\bar{\nabla}$,
$$
\lambda(\bar{\nabla}_{\Xd}D(\Yd,\Zd)+
D(\Xd,\bar{\nabla}_{\Yd}\Zd),\Wd)={\Xd}\tilde D(\Yd,\Zd,\Wd)-
\tilde D(\Yd,\Zd,\bar{\nabla}_{\Xd}\Wd)+\tilde D(\Xd,\bar{\nabla}_{\Yd}\Zd,\Wd)
$$
and
$$
\lambda(D(\Xd,D(\Yd,\Zd)),\Wd)= \tilde D(\Xd,D(\Yd,\Zd),\Wd)=
\lambda(D(\Xd,\Wd),D(\Yd,\Zd)),
$$
this follows from symmetry of the tensor $\tilde D$.
Subtracting the expression with interchanged $\Xd$ and $\Yd$ and
using  symmetry of $\bar{\nabla}$ completes the
proof.
\end{proof}

\begin{coro}\label{coro:flat} Let $g\neq\hat g$ and let the connection
$\nabla^{(g)}$ be flat.
Then the manifold $(\Md,g)$ is conjugate symmetric. Moreover, if $R^{p_0}=0$ for some
$p_0\in (0,1)$ then $R^p=0$ for all $p\in [0,1]$.
\end{coro}

\begin{proof} If $\nabla^{(g)}$ is flat, then also its dual $\nabla^{(\hat g)}$ is
flat, therefore $0=R^1=R^0$ and the manifold is
conjugate symmetric. From Proposition  \ref{prop:Rp}, we see that
$$
0=\bar R(\Xd,\Yd,\Zd,\Wd)+\frac14
\{\lambda(D(\Xd,\Wd),D(\Yd,\Zd)) -\lambda(D(\Xd,\Zd),D(\Yd,\Wd))\}
$$
and therefore
$$
R^p(\Xd,\Yd,\Zd,\Wd)=p(p-1)\{\lambda(D(\Xd,\Wd),D(\Yd,\Zd))-
\lambda(D(\Xd,\Zd),D(\Yd,\Wd))\}
$$
If this vanishes for some $p_0\neq0,1$, then the term in brackets must be zero.
\end{proof}

Let $\lambda$ be the BKM metric and $g(u)=-\log(u)$, then
$\nabla^{(g)}=\nabla^{(m)}$ is flat. It is known
\cite{petz94} that in this case, the metric connection is not flat, hence
$\bar R=R^{1/2}\neq 0$. It follows that $p\nabla^{(m)}+(1-p)\nabla^{(e)}$
is flat if and only if $p=0$ or $p=1$.

\section{Operator calculus.}

In the following sections, we are going to prove that the connection
$\nabla^{(g)}$ is flat if and only if $\nabla^{(g)}=\nabla^{(\alpha)}$ for some
$\alpha\in[-3,3]$. To do this, we will need to
compute the  derivatives of functions of the form $c(L_{\rho},R_{\rho})$.
We use the same method as in \cite{ditt00}.

Let $c$ be a function, defined and complex analytic in a
neighborhood of $(\mathbb R^+)^2$ in $\mathbb C^2$. As the
operators $L_{\rho}$ and $R_{\rho}$ commute and have the same
spectrum as $\rho$, we have by the operator  calculus
$$
c(L_{\rho},R_{\rho})=\frac1{(2\pi i)^2} \int\int
c(\xi,\eta)\frac1{\xi-L_{\rho}}\frac1{\eta-R_{\rho}}d\xi d\eta
$$
where we integrate twice around the spectrum of $\rho$. We have
\begin{eqnarray*}
\frac d{dt}c(L_{\rho+tX},R_{\rho})|_{t=0}&=&\frac1{(2\pi i)^2} \int\int
c(\xi,\eta)\frac1{\xi-L_{\rho}}L_X\frac1{\xi-L_{\rho}}\frac1{\eta-R_{\rho}}d\xi
d\eta\\
\frac{\partial^2}{\partial s\partial
t}c(L_{\rho+sX+tY},R_{\rho})|_{s,t=0}&=&\frac1{(2\pi i)^2} \int\int
c(\xi,\eta)\{
\frac1{\xi-L_{\rho}}L_Y\frac1{\xi-L_{\rho}}L_X\frac1{\xi-L_{\rho}}+\\
&+&\frac1{\xi-L_{\rho}}L_X\frac1{\xi-L_{\rho}}L_Y\frac1{\xi-L_{\rho}}\}
\frac1{\eta-R_{\rho}}d\xi d\eta\\
\frac{\partial^2}{\partial s\partial
t}c(L_{\rho+sX},R_{\rho+tY})|_{s,t=0}&=&
\frac1{(2\pi i)^2}\int\int
c(\xi,\eta)\frac1{\xi-L_{\rho}}L_X\frac1{\xi-L_{\rho}}\frac1{\eta-R_{\rho}}R_Y
\frac1{\eta-R_{\rho}}d\xi d\eta
\end{eqnarray*}

We express the derivatives in form of divided differences, \cite{bhatia}.
Let us denote
\begin{eqnarray}
\label{eq:T1}&T(x,y|z)=\frac{c(x,z)-c(y,z)}{x-y}\\
&T(z|x,y)=\frac{c(z,x)-c(z,y)}{x-y}\\
&T(x,y,z|w)=\frac{T(x,y|w)-T(y,z|w)}{x-z}\\
\label{eq:Tend}&T(x,y|z,w)=\frac{T(x,y|z)-T(x,y|w)}{z-w}=
\frac{T(x|z,w)-T(y|z,w)}{x-y}
\end{eqnarray}
Then we have
\begin{enumerate}
\item[(i)] $T(x,y|z)$, $T(z|x,y)$, $T(x,y|z,w)$ are symmetric in
$x,y$ and $z,w$. $T(x,y,z|w)$ is symmetric in $x,y,z$.
\item[(ii)]
$T(x,x|z)=\frac \partial {\partial x}c(x,z)$ and
$T(z|x,x)=\frac \partial {\partial x}c(z,x)$,
\item[(iii)] $T(x,x,z|w)=\frac \partial {\partial x} T(x,z|w)$,
\item[(iv)] $T(x,x,x|w)=\frac12 \frac{\partial ^2}{\partial x^2}c(x,w)$.
\end{enumerate}

Let $\rho=\sum_i \lambda_i |\psi_i\rangle \langle \psi_i|$ be the spectral
decomposition of $\rho$. Let $e_{ij}=|\psi_i\rangle \langle \psi_j|$, then
$\{e_{ij}\ |\ i,j=1,\dots,n\}$ is a basis of $\Md_n(\mathbb C)$. Let
$u_{ij}=L_{e_{ij}}$, $v_{ij}=R_{e_{ji}}$. Then
$u_{ij}e_{kl}=\delta_{jk}e_{il}$ and $v_{ij}e_{kl}=\delta_{jl}e_{ki}$. We also
have
$$
L_{\rho}=\sum_i\lambda_iu_{ii},\ R_{\rho}=\sum_i\lambda_iv_{ii},\
c(L_{\rho},R_{\rho})=\sum_{i,j}c(\lambda_i,\lambda_j)u_{ii}v_{jj}
$$
Let $X=\sum_{i,j}x_{ij}e_{ij}$.
Inserting this into the expressions for derivatives, we get
\begin{equation}\label{eq:derc}
\frac
d{dt}c(L_{\rho+tX},R_{\rho})|_{t=0}=\sum_{i,j,k}T(\lambda_i,\lambda_j|\lambda_k)x_{ij}u_{ij}v_{kk}
\end{equation}
Similarly,
\begin{eqnarray}
\frac{\partial^2}{\partial s\partial
t}c(L_{\rho+sX+tY},R_{\rho})|_{s,t=0}&=&
\sum_{i,j,k,l}T(\lambda_i,\lambda_j\lambda_k|\lambda_l)(x_{ij}y_{jk}+y_{ij}x_{jk})u_{ik}v_{ll}\\
\frac{\partial^2}{\partial s\partial
t}c(L_{\rho+sX},R_{\rho+tY})|_{s,t=0}&=&\sum_{i,j,k,l}T(\lambda_i,\lambda_j|
\lambda_k,\lambda_l)x_{ij}y_{lk}u_{ij}v_{kl}
\end{eqnarray}

\section{Conjugate symmetry.}

Let $g\in G$ and let $k$ be given by (\ref{eq:k}). Let us define
the function $c:\ \mathbb R^+\times \mathbb R^+\to \mathbb R$ by
$c(x,y)=1/yk(x/y)$. As we see from the integral representation,
the function $k$ is operator monotone decreasing, therefore it has
an analytic extension to the right halfplane in $\mathbb C$. It
follows that $c$ is complex analytic in a neighborhood of
$(\mathbb R^+)^2$ and we may use the results of the previous
section. Note also that for $\hat g$, $\hat k(u)=u^{-1}k(u^{-1})$
and $\hat c(x,y)=c(y,x)$. Moreover, $\bar c(x,y)=c(x,y)+  \hat
c(x,y)=1/yk_{sym}(x/y)$ is the Chentsov-- Morozova function.  As
follows from (\ref{eq:k}), $k(1)=\int_{[0,\infty]}d\mu =1/2$, therefore $c(x,x)=\frac1{2x}$ for all $g$.

\begin{lemma}\label{lemma:coeff} Let $c$ and $\hat c$ be as above. Then
\begin{eqnarray*} \lambda_{\rho}(\nabla^{(g)}_{\Xd}\Yd,\Zd)=
2{\rm Re}\frac d{ds}\Tr\{X\hat c(L_{\rho+sY},R_{\rho})(Z)+Y\hat
c(L_{\rho+sX},R_{\rho})(Z)-\\- Xc(L_{\rho+sZ},R_{\rho})(Y)\}|_{s=0}
\end{eqnarray*}

 \end{lemma}

\begin{proof} From Theorem \ref{theorem:Hg} we compute
\begin{eqnarray*}
\lambda_{\rho}(\nabla^{(g)}_{\Xd}\Yd,\Zd)&=&-\frac{\partial^3}
{\partial s\partial t\partial
u}(uZ-sX-tY)c(L_{\rho+uZ},R_{\rho+sX+tY})(uZ-sX-tY)|_{s,t,u=0}=\\
&=&-\frac d{ds} \Tr
\{Xc(L_{\rho+sZ},R_{\rho})(Y)+Yc(L_{\rho+sZ},R_{\rho})(X)-\\
&-& X c(L_{\rho},R_{\rho+sY})(Z)-Z
c(L_{\rho},R_{\rho+sY})(X)-\\
&-& Y c(L_{\rho},R_{\rho+sX})(Z)-Z
c(L_{\rho},R_{\rho+sX})(Y)\}|_{s=0}
\end{eqnarray*}

For $\sigma,\rho\in \Md$, $c(L_{\sigma},R_{\rho})$ is a positive operator on
$\Md_n(\mathbb C)$ endowed with the inner product
$\langle A,B\rangle =\Tr A^*B$. For hermitian $X$ and $Y$, we have
$$
\Tr Xc(L_{\sigma},R_{\rho})(Y)+\Tr Yc(L_{\sigma},R_{\rho})(X)=
2{\rm Re}\Tr Xc(L_{\sigma},R_{\rho})(Y)
$$
Clearly, for all $X\in \Md_h$ and sufficiently small $s$, $\rho+sX\in \Md$.
Moreover,
$$
{\rm Re}\ \Tr Xc(L_{\rho},R_{\rho+sY})(Z)={\rm Re}\ \Tr
(Xc(L_{\rho},R_{\rho+sY})(Z))^*={\rm Re}\ \Tr X\hat
c(L_{\rho+sY},R_{\rho})(Z)
$$
\end{proof}

\begin{lemma}\label{lemma:tilD} Let $D(\Xd,\Yd)=
\nabla^{(g)}_{\Xd}\Yd-\nabla^{(\hat g)}_{\Xd}\Yd$ and let $\tilde
D(\Xd,\Yd,\Zd)=\lambda(D(\Xd,\Yd),\Zd)$. Let us denote $c_r(x,y)=\hat
c(x,y)-c(x,y)=c(y,x)-c(x,y)$ and let
$$
Q(X,Y,Z)=\frac d{ds}\Tr Xc_r(L_{\rho+sY},R_{\rho})(Z)
$$
Then
$$
\tilde D(\Xd,\Yd,\Zd)=2{\rm Re}\{ Q(X,Y,Z)+Q(Y,X,Z)+Q(X,Z,Y)\}=6Q_{sym}(X,Y,Z)
$$
where $Q_{sym}$ is the symmetrization of $Q$ over $X$, $Y$, $Z$.
\end{lemma}

\begin{proof} Straightforward from previous Lemma.\end{proof}

Let us now denote by $\bar T(x,y|z)$ resp. $R(x,y|z)$, etc. the expressions
(\ref{eq:T1})...(\ref{eq:Tend}) for $c=\bar c$ resp. $c=c_r$.
Using the previous section, we find
\begin{equation}\label{eq:Q}
Q(X,Y,Z)=\sum_{i,j,k}R(\lambda_i,\lambda_j|\lambda_k)x_{ki}y_{ij}z_{jk}
\end{equation}
Further,
\begin{eqnarray}\label{eq:XQ}
\Xd Q(Y,Z,W)&=&\sum_{i,j,k,l}R(\lambda_i,\lambda_j,\lambda_k|\lambda_l)
(x_{ij}z_{jk}+z_{ij}x_{jk})w_{kl}y_{li}+\\
&+&\sum_{i,j,k,l}R(\lambda_i,\lambda_j|\lambda_k,\lambda_l)
z_{ij}w_{jl}x_{lk}y_{ki}\nonumber
\end{eqnarray}
Clearly, $\Xd\tilde D(\Yd,\Zd,\Wd)$ is the symmetrization of (\ref{eq:XQ})
over $Y,Z,W$.

\begin{prop}\label{prop:metric.connection}
Let
$$
S(x,y|z)=\frac1{2\bar c(x,y)}\{\bar T(x,z|y)+\bar T(y,z|x)-\bar T(x,y|z)\}
$$
Then the -1-representation
$\bar{\nabla}_{\Xd}\Yd(\rho)=
\sum_{\alpha,\beta}d_{\alpha\beta}e_{\alpha\beta}$
where
$$
d_{\alpha\beta}=\sum_iS(\lambda_{\alpha},\lambda_{\beta}|\lambda_i)(x_{\alpha i}y_{i\beta}+y_{\alpha i}x_{i\beta})
$$
\end{prop}

\begin{proof} Let $h=\frac12(g+\hat g)$, then $\bar{\nabla}=\nabla^{(h)}$. In this
case, $c=\frac12\bar c=\hat c$.
From Lemma \ref{lemma:coeff} and (\ref{eq:derc}), we see that
\begin{equation}\label{eq:coeff.metric}
\lambda_{\rho}(\bar{\nabla}_{\Xd}\Yd(\rho),Z)={\rm Re}\sum_{i,j,k}\bar
T(\lambda_i,\lambda_j|\lambda_k)\{x_{ki}y_{ij}z_{jk}+y_{ki}x_{ij}z_{jk}-x_{ki}z_{ij}y_{jk}\}
\end{equation}
Let us denote $f^1_{\alpha\alpha}=e_{\alpha\alpha}$, for $\alpha=1,\dots n$,
$f^2_{\alpha\beta}=e_{\alpha\beta}+e_{\beta\alpha}$, $\alpha\neq\beta$ and
$f^3_{\alpha\beta}=i(e_{\alpha\beta}-e_{\beta\alpha})$, $\alpha\neq\beta$. Then
$\{f^1_{\alpha\alpha}, \alpha=1,\dots,n, f^k_{\alpha\beta}, \ k=2,3,\alpha<\beta=2,\dots,n\}$ forms a
basis of $T_{\rho}$ with elements mutually orthogonal with respect to each
monotone metric $\lambda$. Moreover,
$$
\lambda(f^k_{\alpha\beta},f^k_{\alpha\beta})=
\left\{ \begin{array}{lc}
\bar c(\lambda_{\alpha},\lambda_{\alpha})& k=1\\
2\bar c(\lambda_{\alpha},\lambda_{\beta})& k\neq 1\\
\end{array}\right.
$$
Suppose that
$$
\bar{\nabla}_{\Xd}\Yd(\rho)=\sum_{k,\alpha\leq \beta}a^k_{\alpha\beta}f^k_{\alpha\beta},
$$
then $\bar{\nabla}_{\Xd}\Yd(\rho)=\sum_{\alpha,\beta}d_{\alpha\beta}e_{\alpha\beta}$, where
$d_{\alpha\alpha}=a^1_{\alpha\alpha}$,
$d_{\alpha\beta}=a^2_{\alpha\beta}+ia^3_{\alpha\beta}$, if $\alpha< \beta$ and $d_{\alpha\beta}=a^2_{\alpha\beta}-ia^3_{\alpha\beta}$, if
$\alpha>\beta$.
From (\ref{eq:coeff.metric}) we compute
\begin{eqnarray*}
a^1_{\alpha\alpha}&=&2{\rm
Re}\sum_jS(\lambda_{\alpha},\lambda_{\alpha}|\lambda_j)x_{\alpha
j}y_{j\alpha}\\
a^2_{\alpha\beta}&=&{\rm Re}\sum_jS(\lambda_{\alpha},\lambda_{\beta}|\lambda_j)
\{x_{\alpha j}y_{j\beta}+y_{\alpha j}x_{j\beta}\}\\
a^3_{\alpha\beta}&=&{\rm Im}\sum_jS(\lambda_{\alpha},\lambda_{\beta}|\lambda_j)
\{x_{\alpha j}y_{j\beta}+y_{\alpha j}x_{j\beta}\}
\end{eqnarray*}
\end{proof}

As we know from section \ref{section:pconnections},
$(\Md,g)$ is conjugate symmetric if and only if
\begin{eqnarray}\label{eq:conjugateD}
&\Xd\tilde D(\Yd,\Zd,\Wd)-\Yd\tilde D(\Xd,\Zd,\Wd)+
\tilde D(\Xd,\bar{\nabla}_{\Yd}\Zd,\Wd)+\tilde
D(\Xd,\Zd,\bar{\nabla}_{\Yd}\Wd)\\
&-\tilde
D(\Yd,\bar{\nabla}_{\Xd}\Zd,\Wd)-\tilde D(\Yd,\Zd,\bar{\nabla}_{\Xd}\Wd)
=0\nonumber
\end{eqnarray}

Using Lemma \ref{lemma:tilD}, (\ref{eq:Q}), (\ref{eq:XQ}) and Proposition
\ref{prop:metric.connection}, we express the above equality in terms of the
divided differences and then insert the basis elements $f^k_{\alpha\beta}$.
This, and other further lengthy computations, is best performed using some
software suitable for symbolic calculations, like Maple or Mathematica.

The equalities
$\bar c(x,y)=\bar c(y,x)$, $c_r(x,y)=-c_r(y,x)$ and the definition and
properties of divided differences imply that
\begin{eqnarray}
\label{eq:RTS1}R(x,y|x)&=&\frac1{x-y}c_r(x,y)=R(x,y|y)\\
R(x,x|x)&=&-\frac{\alpha}{6x^2}, \mbox{ where } \alpha=2g'''(1)+3\\
R(x,y|z,w)&=&-R(z,w|x,y)\\
\bar T(x,y|z,w)&=&\bar T(z,w|x,y)\\
S(x,y|x)&=&\frac 12 \frac \partial {\partial x}\log\bar c(x,y)\\
S(x,x|y)&=&\frac 12\{2\frac{1-x\bar c(x,y)}{x-y}-x\frac \partial {\partial x}
\bar c(x,y)\}\\
\label{eq:RTSend}S(x,x|x)&=&-\frac 1{4x}
\end{eqnarray}
for all $x,y,z,w>0$.

\begin{theorem}\label{theorem:conjugate} Let $g\neq \hat g$ and let $\bar g=g+\hat g$, $g_r=\hat g-g$.
 If $(\Md,g)$ is conjugate symmetric, then
\begin{equation}\label{eq:conjugate}
-\alpha \bar g(u)=2ug_r'(u)-g_r(u)+2au+2a
\end{equation}
for all $u>0$, here $a=g'(1)$ and $\alpha=2g'''(1)+3$.
\end{theorem}

\begin{proof} Let us write the equality (\ref{eq:conjugateD}) for
the basis elements $f^k_{\alpha\beta}$ with $\alpha,\beta\in\{1,2\}$,
in this case, the
resulting expression depends only from eigenvalues $\lambda_1$ and
$\lambda_2$ of $\rho$. Let us put $X=Z=e_{11}$ and
$Y=W=e_{12}+e_{21}$ and let $\lambda_1=x$, $\lambda_2=y$. We get
\begin{eqnarray*}
&R(x,x,x|y)-R(x,x,y|x)+R(x,y|x,x)-R(x,x|x,y)+3R(x,x|x)S(x,x|y)-\\
&S(x,x|x)(2R(x,y|x)+R(x,x|y))=0
\end{eqnarray*}
We have
\begin{eqnarray*}
c_r(x,y)&=&\frac{yg_r(x/y)}{(x-y)^2}+\frac{2a}{x-y},\\
\bar c(x,y)&=&\frac{y\bar g(x/y)}{(x-y)^2}
\end{eqnarray*}
From this and from (i)...(iv),
(\ref{eq:RTS1})...(\ref{eq:RTSend}), we get the equation
$$
2g''_r(\frac xy)\frac xy+2a+\alpha\bar g'(\frac xy)
+g_r'(\frac xy)=0
$$
Putting $u=x/y$ and integrating this, taking into account that $\bar g(1)=0$,
$g_r(1)=0$ and $g_r'(1)=-2a$, we get (\ref{eq:conjugate}).

\end{proof}

\begin{remark} Let $g\neq \hat g$, $\alpha$ and $a$ be as above.
Then according to the previous theorem,
if $(\Md,g)$ is conjugate symmetric, then
\begin{equation}\label{eq:conjugate2}
\frac{1+\alpha}2\bar g(u)=g'(u^{-1})+ug'(u)-au-a
\end{equation}

If $h$ is symmetric, then $(\Md,h)$ is, of course, conjugate symmetric. In such
a case, $\alpha=a=0$ and the
equation (\ref{eq:conjugate2}) reads
$$
h(u)=h'(u^{-1})+uh'(u)
$$
which is fulfilled for all symmetric $h\in G$.
\end{remark}

\begin{example} It is easily checked that (\ref{eq:conjugate2}) is satisfied for
all $pg_{\alpha}+(1-p)g_{-\alpha}$, $p\in[0,1]$, $\alpha\in [-3,3]$
(as it should be). On the other hand, it is
not true for $g_s$ from the extreme boundary of $G$, unless $s=1$, which is
symmetric (the Bures case),  or $s=0$, which corresponds to $g_{\alpha}$,
$\alpha=3$.
\end{example}

\section{Flat connections.}

As we know from Corollary \ref{coro:flat} and Proposition
\ref{prop:Rp}, the connection $\nabla^{(g)}$ is flat if
and only if
\begin{enumerate}
\item[(a)] $(\Md,g)$ is conjugate symmetric
\item[(b)] $\bar R(\Xd,\Yd,\Zd,\Wd)+\frac14\{
\lambda(D(\Xd,\Wd),D(\Yd,\Zd))-\lambda(D(\Xd,\Zd),D(\Yd,\Wd))\}=0$
\end{enumerate}
This holds also for symmetric $g$, in that case (a) is satisfied and $D=0$.

\begin{lemma}\label{lemma:barR}
$$
\bar R(\Xd,\Yd,\Zd,\Wd)=\Xd\lambda(\bar {\nabla}_{\Yd}\Zd,\Wd)-
\Yd\lambda(\bar {\nabla}_{\Xd}\Zd,\Wd)+
\lambda(\bar{\nabla}_{\Xd}\Zd,\bar{\nabla}_{\Yd}\Wd)-
\lambda(\bar{\nabla}_{\Yd}\Zd,\bar{\nabla}_{\Xd}\Wd)
$$
\end{lemma}

\begin{proof} The statement is proved similarly as Proposition \ref{prop:Rp},
using self-duality and symmetry of $\bar{\nabla}$.
\end{proof}

As before, we compute
$$
\Xd\lambda(\bar{\nabla}_{\Yd}\Zd,\Wd)=
{\rm Re}\{\Xd\bar Q(Y,Z,W)+\Xd\bar Q(Z,Y,W)-
\Xd\bar Q(Y,W,Z)\}
$$
where
\begin{eqnarray}\label{eq:XbarQ}
\Xd\bar Q(Y,Z,W)&=&\sum_{i,j,k,l}\bar T(\lambda_i,\lambda_j,\lambda_k|\lambda_l)
(x_{ij}z_{jk}+z_{ij}x_{jk})w_{kl}y_{li}+\\
&+&\sum_{i,j,k,l}\bar T(\lambda_i,\lambda_j|\lambda_k,\lambda_l)
z_{ij}w_{jl}x_{lk}y_{ki}\nonumber
\end{eqnarray}

Moreover,
\begin{equation}\label{eq:lambda}
\lambda(X,Y)=\Tr X\bar c(L_{\rho},R_{\rho})(Y)=\sum_{i,j}\bar
c(\lambda_i,\lambda_j)x_{ji}y_{ij}
\end{equation}

The second term in (b) can be written in a form using $\tilde D$:
let $\{b_j,\ j=1,\dots N\}$ be the orthonormal basis obtained by
normalization of
$\{f^k_{\alpha\beta},k=1,2,3, \alpha\leq \beta =1,\dots,n\}$, then
\begin{eqnarray}\label{eq:second.term}
&\lambda(D(\Xd,\Wd),D(\Yd,\Zd))-\lambda(D(\Xd,\Zd),D(\Yd,\Wd))=\\
&\sum_j\{\tilde
D(X,W,b_j)\tilde D(Y,Z,b_j)-\tilde D(X,Z,b_j)\tilde D(Y,W,b_j)\}\nonumber
\end{eqnarray}

Using Lemma \ref{lemma:barR}, (\ref{eq:XbarQ}), (\ref{eq:lambda}),
(\ref{eq:second.term}) and Proposition \ref{prop:metric.connection},
we get from (b) an equation involving divided
differences, and we may proceed the same way as in the last section.

\begin{prop}\label{prop:flat} Let $g\in G$. If the connection $\nabla^{(g)}$ is flat, then
\begin{equation}\label{eq:flat}
(\alpha^2-1)\bar g(u) +\bar g'(u)(u-1)-2\bar
g''(u)u(1+u)+\alpha(g_r'(u)+2a)(u-1)+8=0
\end{equation}
for all $u>0$.
\end{prop}

\begin{proof}
Let $X=Z=e_{11}$ and $Y=W=e_{12}+e_{21}$. From (b), we get the equation
\begin{eqnarray*}
&2\bar T(x,x,x|y)-2\bar T(x,x|x,y)-2\bar c(x,y)S(x,y|x)^2+4\bar
c(x,x)S(x,x|x)\bar S(x,x|y)-\\
&-3\frac{R(x,x|x)}{\bar c(x,x)}(2R(x,y|x)+R(x,x|y))+
\frac1{\bar c(x,y)}(2R(x,y|x)+R(x,x|y))^2=0
\end{eqnarray*}
For  $X=Z=e_{12}+e_{21}$, $Y=W=i(e_{12}-e_{21})$, the equation (b) reads
\begin{eqnarray*}
&4\bar T(y,y,x|x)+4\bar T(x,x,y|y)-8\bar T(x,y|x,y)+4\bar c(x,x)S(x,x|y)^2+
4\bar c(y,y)S(y,y|x)^2-\\
&-(2R(x,y|x)+R(x,x|y))^2-(2R(x,y|y)+R(y,y|x))^2=0
\end{eqnarray*}
As in the proof of Theorem \ref{theorem:conjugate}, we get after some
rearrangements
$$
u\left[ (g_r'(u)+2a)^2-(\bar g'(u))^2\right]+\bar g(u)\{ 2u\bar g''(u)+\bar
g'(u)+\alpha(g_r'(u)+2a)\}=0
$$
from the first equation and
$$
u\left[(g_r'(u)+2a)^2-(\bar g'(u))^2\right]+
\{g_r'(u)u-g_r(u)+2a\}^2-
\{\bar g'(u)u-\bar g(u)\}^2+8\bar g(u)=0
$$
from the second equation.

If $g$ is symmetric, then in the above two equations
$\alpha=a=0$ and $g_r=0$. From this we get
$$
\bar g(u)\{-\bar g(u) +\bar g'(u)(u-1)-2\bar
g''(u)u(1+u)+8\}=0
$$
which is (\ref{eq:flat}).

Let now $g\neq \hat g$. From (a), $(\Md, g)$ is conjugate symmetric,
and therefore (\ref{eq:conjugate}) holds. From this
$$
g_r'(u)u-g_r(u)+2a=-\alpha\bar g(u)-u\{g_r'(u)+2a\}
$$
Inserting this into the second equation and after some further computation,
we get (\ref{eq:flat}).

\end{proof}

We are now in position to prove our main theorem.
\begin{theorem} \label{theorem:flatg}
Let $g\in G$ and $\alpha=2g'''(1)+3$. Then $\alpha\in [-3,3]$ and
the connection $\nabla^{(g)}$ is flat if and only if $\nabla^{(g)}=\nabla^{(\alpha)}$.
\end{theorem}

\begin{proof}
Let $g$ be symmetric and suppose that $\nabla^{(g)}$ is
flat. Then $\bar g=2g$ and we get from (\ref{eq:flat}) that $g$
is the solution of
$$
-g(u) + g'(u)(u-1)-2 g''(u)u(1+u)+4=0
$$
with initial conditions $ g(1)=0$, $g'(1)=0$. The unique solution of
this equation is
$$
g(u)=2(1-\sqrt u)^2=g_0
$$

If $g\ne \hat g$, then from (\ref{eq:conjugate}) and (\ref{eq:flat}) we get
that $g_r$ is the solution of
\begin{eqnarray*}
&(\alpha^2-1)g_r(u)-(\alpha^2-1)(1+u)g_r'(u)+4u(u+2)g_r''(u)+\\
&+4u^2(u+1)g_r'''(u)-4a(\alpha^2-1)+8\alpha=0
\end{eqnarray*}
with $g_r(1)=0$, $g_r'(1)=-2a$ and $g_r''(1)=0$. If $\alpha\neq
\pm1$, the unique solution is
\begin{eqnarray*}
g_r(u)&=&\frac4{1-\alpha^2}(u^{\frac{1+\alpha}2}-u^{\frac{1-\alpha}2})-(\frac
{4\alpha}{1-\alpha^2}+2a)(u-1)=\\
&=&g_{-\alpha}(u)-g_{\alpha}(u)-2(a-g'_{\alpha}(1))(u-1)
\end{eqnarray*}
and from (\ref{eq:conjugate}), we get $\bar g=g_{\alpha}+g_{-\alpha}$.

If $\alpha=-1$, then the solution of the above equation is
$$
g_r(u)=\log(u)(u+1)-2(a-g'_{-1}(1))(u-1)
$$
and from (\ref{eq:conjugate}) we get
$$
 \bar g(u)=\log(u)(u-1)
$$
It follows that $g=g_{\alpha}$, up to an additional linear term
$(g'(1)-g'_{\alpha}(1))(u-1)$.
\end{proof}

\begin{coro} Let $\lambda$ be a monotone Riemannian metric and let
$\bar{\nabla}$ be the metric connection. Then $\bar{\nabla}$ is flat if and
only if $\lambda$ is the WY metric ($\alpha=0$).
\end{coro}

\begin{proof} Let $G_h$ be the convex subset of $G$, corresponding to
$\lambda$. Then $\bar{\nabla}=\nabla^{(h)}$ and $h=\hat h$ implies that
$h'''(1)=-\frac 32$. The proof now follows from
Theorem \ref{theorem:flatg}.
\end{proof}

\end{document}